\documentclass[12pt]{iopart}
\usepackage{iopams}

\newcommand{\DIV}[1][\normalsize]{\,\mbox{div}\,}
\newcommand{\INT}[3]{\int_{#1}^{#2} \, \rmd#3 \,}
\newcommand{\SP}[2]{\left\langle #1, #2 \right\rangle}
\newcommand{\wt}[1]{\widetilde{#1} }
\newcommand{\abs}[1]{\left| #1 \right|}

\def\N{ {\mathbb N} }
\def\R{ {\mathbb R} }
\def\C{ {\mathbb C} }

\begin{document}

\title[Just dust]{Just dust : About the (in)applicability of rotating dust solutions as realistic galaxy models}
\author{Tobias Zingg\footnote{Email: Tobias.Zingg@stud.unibas.ch},
Andreas Aste and Dirk Trautmann}
\address{Department for Physics and Astronomy, University of Basel,\\
4056 Basel, Switzerland}
\date{August, 2006}
\begin{center}
\begin{abstract}
\noindent
Solutions of the stationary axisymmetric Einstein equations describing the interior of
circularly rotating dust are investigated in order to study their potential applicability
as galaxy models.
It is shown that such interior solutions cannot be used as global metrics without
becoming unphysical in certain regions of space.
Although definite results concerning the non-existence of a continuation into a vacuum can only be derived for interior solutions describing rigidly rotating dust, the present analysis exhibits that the case of non-rigidly rotating dust would in general also be inadequate as a physically reasonable galaxy model.
\end{abstract}
\end{center}
\vskip 0.1 cm

\section{Introduction}

There has been some recent interest in a general relativistic model of
Cooperstock \& Tieu \cite{2005astro.ph..7619C}, which attempts to fit rotation curves
of spiral galaxies without invoking dark matter.
This model is constructed such that it gives an excellent fit
to observed rotation curves. However, there have been serious concerns about
the consistency of the model in general
\cite{2006astro.ph..3791B,2006astro.ph..1191C}.

By using some basic results presented in \cite{1975JMP....16.1806W}, where solutions
of the Einstein equations describing axisymmetric rotating dust were calculated,
the applicability of the corresponding solutions to serve as galaxy models is studied
in this paper.
Such axisymmetric solutions have been used as toy models
for galaxies during the last years, e.g. in \cite{2005astro.ph..7619C}
and \cite{2006astro.ph..2519B}, in order to investigate whether general relativity can
give new insights into the problem of the distribution of dark matter in our universe.

The crucial question concerning the usefulness of axisymmetric dust solutions
is whether the assumption of galaxies being stationary axisymmetric objects is in some way
realistic.
Since a galaxy is a dynamic object and a spiral structure is surely not invariant
under arbitrary rotations, the above seems barely reasonable.
Furthermore, circular flow of dust makes only sense for the galactic disk,
but far off the galactic plane such dynamics would become rather unphysical.
This has already been pointed out in \cite{2006astro.ph..1191C}, where it was also shown,
that the models in \cite{2005astro.ph..7619C,2006astro.ph..2519B} indeed
describe rigid rotation.
Nevertheless, for a slowly rotating galactic disk where details of the spiral structure can be
'washed' out, i.e. where variations in the time coordinate $t$ and the angular
coordinate $\varphi$ are rather small in comparison to variations in the other
coordinates, the assumptions described above might yield useful approximations.
A restriction to the galactic disk may also be reasonable in connection with the debate about
dark matter, since the main question is then whether an inner galactic disk solution can be
continued into a vacuum without the need for a dark matter halo.
Indeed it will be show below that solutions describing simple axisymmetric configurations
of rigidly rotating dust cannot be continued into a vacuum solution without
involving singularities.

Furthermore, also axisymmetric dust solutions with non-rigid rotation will be considered.
In comparison to rigid rotation models, these solutions have a further degree of freedom
expressible by an arbitrary function of one variable.
A general answer about transitions into vacuum solutions will not be given, but according
to what is currently known about vacuum solutions for Dirichlet boundary data
\cite{1998CMaPh.190..509S,2002PhRvD..65d4006A,1998MPLA...13.1509M},
such transitions will in general exhibit singular behavior on the boundary.

\section{Einstein equations for static axisymmetric metrics}

Throughout this paper, geometric units are used such that $8\pi G = c = 1$.
In the case of a static axisymmetric problem as treated in the following,
the metric can be assumed to have the simplified form
\begin{equation}  \label{EQ:metric}
g = \rme^{2W}(\rmd t - N\,\rmd \varphi)^2 - \rme^{-2W}\left[R^2\,\rmd \varphi^2 +
\rme^{2V}(\rmd \mu^2 + \rmd \nu^2) \right] \, ,
\end{equation}
where the functions $W,V,R$ and $N$ depend only on the coordinates $\mu$ and $\nu$.
Details about this simplification can be found in, e.g. \cite{Wald:GR}.
The form of the metric \eref{EQ:metric} includes the assumption that $\partial_t$ is
timelike and that the planes orthogonal to $\partial_t$ and $\partial_\varphi$, i.e.
$\partial_\mu$ and $\partial_\nu$, are spacelike.
This obviously excludes the possibility that the metric
can have highly gravitating domains like ergoregions.

In this paper, calculations are performed in the tetrad
\begin{equation}\fl  \label{EQ:E-vecs}
E_0 = \rme^{-W} \partial_t\, , \qquad
E_1 = \rme^{W-V} \partial_\mu\, , \qquad
E_2 = \rme^{W-V} \partial_\nu\, , \qquad
E_3 = \frac{\rme^{W}}{R}[N\partial_t+\partial_\varphi] \, ,
\end{equation}
where $g(E_i,E_j)$ has the form of the Minkowski tensor. This semi-orthogonal
basis is used since it allows some simplifications in the derivation
of the curvature forms and the Einstein equations (see also \cite{Oloff:GdR}).
However, the only apparent difference to the coordinate basis is a slightly different
representation of the used tensors.

\paragraph{The stress energy tensor}$ $

The energy tensor of circularly rotating dust (i.e. a pressureless fluid) is defined via
\begin{equation}\label{EQ:matter}
T(x,y) = \rho\,g(x,Z)g(y,Z) \, ,
\end{equation}
where the timelike $4$-velocity of the fluid $Z$ is given by
\begin{equation}  \label{EQ:matter_Z}
Z = \cosh \frac{\Lambda}{2} \, E_0 - \sinh \frac{\Lambda}{2} \, E_3
\end{equation}
and $\rho$, as well as $\Lambda$, depend only on $\mu$ and $\nu$.
Here, $\Lambda$ has been introduced as a degree of freedom for the $4$-velocity $Z$,
but if $E_0$ and $R E_3$ are interpreted as some kind of time and angular
vector, respectively, then the value
\begin{equation}  \label{EQ:X}
X = \frac{1}{R} \tanh \frac{\Lambda}{2}
\end{equation}
can be interpreted as the angular velocity of the dust cloud in the frame \eref{EQ:E-vecs}.

Furthermore, when writing $Z$ as parallel to
$\partial_t-\Omega\partial_\varphi$, where $\Omega$ is the angular velocity
seen by an observer who is not rotating with respect to the background of
fixed stars, $\Omega$ and $X$ are related via
\begin{equation}  \label{EQ:angovelo}
X = \frac{\Omega}{\rme^{2W}(1+\Omega N)}  \, ,
\end{equation}
which can also be written as
\begin{equation}  \label{EQ:X_angovelo}
\Omega = \frac{X}{\rme^{-2W} - XN}\, .
\end{equation}
Now, the condition, that $\partial_t-\Omega\partial_\varphi$ must be timelike,
which is equivalent to demand $Z$ being timelike, can be written as
\begin{equation}  \label{EQ:rot_cond}
\Omega^2 R^2 < (1+\Omega N)^2 \rme^{4W}  \, .
\end{equation}
According to \eref{EQ:angovelo}, this condition is equivalent to $\abs{RX}<1$.
This inequality has the relatively simple physical interpretation, namely
that the velocity $XR$ of a particle at radius $R$ in the frame \eref{EQ:E-vecs}
cannot exceed the speed of light.

As a consequence of Bianchi's identity, the tensor $T$ must obey the condition $\DIV T = 0$.
Since $\rho$ only depends on the coordinates $\mu$ and $\nu$, this is equivalent to the equations
\numparts \begin{eqnarray}\fl
\rho\, \DIV Z &= 0\label{EQ:divfree.a}
\\\fl
\rho\, \nabla_{Z}Z  &= 0 \, , \label{EQ:divfree.b}
\end{eqnarray} \endnumparts
where $\nabla$ denotes the Levi-Civita connection on the considered Lorentzian
manifold.
Due to the form of the metric, \eref{EQ:divfree.a} is automatically fulfilled,
whereas \eref{EQ:divfree.b} implies the further conditions
\numparts \begin{eqnarray}\fl
\rho \rme^{W-V}\left[ W_\mu \cosh\Lambda + \rme^{2W}\frac{N_\mu}{2R}
\sinh\Lambda - (\cosh\Lambda-1) \frac{R_\mu}{2R} \right] &= 0
\label{EQ:divfree2.a}
\\\fl
\rho \rme^{W-V}\left[ W_\nu \cosh\Lambda + \rme^{2W}\frac{N_\nu}{2R}
\sinh\Lambda - (\cosh\Lambda-1) \frac{R_\nu}{2R} \right] &= 0  \, .\label{EQ:divfree2.b}
\end{eqnarray} \endnumparts
The indices denote partial derivatives.
In the regions where $\rho\neq 0$, it will be advantageous to use these equations
instead of a subset of the Einstein equations.

\paragraph{Einstein equations}$ $

With the stress energy tensor according to \eref{EQ:matter}, the Einstein equations can be
represented as
\numparts \begin{equation}\fl
\rme^{2(W-V)} \frac{R_{\mu\mu}-R_{\nu\nu}}{R} = 0\label{EQ:Einstein.a}
\end{equation}
\begin{equation}\fl
\rme^{2(W-V)} \left[ 2W_{\mu\mu} + 2W_{\nu\nu} + 2\frac{R_\mu W_\mu}{R}
+ 2\frac{R_\nu W_\nu}{R} + \rme^{4W}\frac{N_\mu^2+N_\nu^2}{R^2} \right]
= \rho \cosh\Lambda\label{EQ:Einstein.b}
\end{equation}
\begin{equation}\fl
\rme^{2(W-V)} \frac{\rme^{2W}}{R} \left[ N_{\mu\mu} + N_{\nu\nu} +
4 W_\mu N_\mu + 4 W_\nu N_\nu
  - \frac{R_\mu N_\mu+R_\nu N_\nu}{R} \right] = -\rho \sinh\Lambda
\label{EQ:Einstein.c}
\end{equation}
\begin{equation}\fl
\rme^{2(W-V)} \left[ V_{\mu\mu} + V_{\nu\nu} + W_\mu^2 + W_\nu^2
+ \rme^{4W}\frac{N_\mu^2+N_\nu^2}{4R^2} \right] = \rho\sinh^2\frac{\Lambda}{2}
\label{EQ:Einstein.d}
\end{equation}
\begin{equation}\fl
\rme^{2(W-V)} \left[ W_\mu^2 - W_\nu^2 - \frac{R_\mu V_\mu}{R} +
\frac{R_\nu V_\nu}{R} + \frac{R_{\nu\nu}}{2R}
- \frac{R_{\nu\nu}}{2R} - \rme^{4W}\frac{N_\mu^2-N_\nu^2}{4R^2}
\right] = 0  \label{EQ:Einstein.e}
\end{equation}
\begin{equation}\fl
\rme^{2(W-V)} \left[ 2W_\mu W_\nu - \frac{R_\mu V_\nu}{R} -
\frac{R_\nu V_\mu}{R}
+ \frac{R_{\mu\nu}}{R} - \rme^{4W}\frac{N_\mu N_\nu}{2R^2}
\right] = 0 \, .  \label{EQ:Einstein.f}
\end{equation} \endnumparts
If $R_\mu^2+R_\nu^2 \neq 0$, which makes sense if $R$ is interpreted as some kind of radial
distance from the rotation axis, then (\ref{EQ:Einstein.a},$e$,$f$) can be used to find
explicit expressions for $V_\mu$ and $V_\nu$ as follows
\numparts \begin{equation}\fl
\eqalign{ V_\mu =& \frac{1}{2}\partial_\mu \ln (R_\mu^2+R_\nu^2)
+ \frac{ R_\mu W_\nu^2-R_\mu W_\mu^2-2R_\nu W_\mu W_\nu }{4R(R_\mu^2+R_\nu^2)} \\
& \quad + \rme^{4W}\frac{ R_\mu N_\nu^2-R_\mu N_\mu^2-2R_\nu N_\mu N_\nu }{4R(R_\mu^2+R_\nu^2)} }  \label{EQ:V.a}
\end{equation}
\begin{equation}\fl
\eqalign{ V_\nu =& \frac{1}{2}\partial_\nu \ln (R_\mu^2+R_\nu^2)
+ \frac{ -R_\nu W_\nu^2+R_\nu W_\mu^2-2R_\mu W_\mu W_\nu }{4R(R_\mu^2+R_\nu^2)} \\
  & \quad + \rme^{4W}\frac{ -R_\nu N_\nu^2+R_\nu N_\mu^2-2R_\mu N_\mu N_\nu }{4R(R_\mu^2+R_\nu^2)}\, . }\label{EQ:V.b}
\end{equation} \endnumparts
The integrability condition for (\ref{EQ:V.a},$b$) is a consequence of (\ref{EQ:divfree2.a},$b$) and (\ref{EQ:Einstein.a}-$c$).
Furthermore, \eref{EQ:Einstein.d} can be derived from those equations.
Thus, the original problem of solving (\ref{EQ:Einstein.a}-$f$) can be reduced to finding solutions of
(\ref{EQ:divfree2.a},$b$) and (\ref{EQ:Einstein.a}-$c$).

By introducing the flat pseudo-metric
\begin{equation}\label{EQ:pseudo_metric}
g^{\cal P} = d\mu^2 + d\nu^2
\end{equation}
and defining the derivative operators $\DIV^{\cal P}$, $\nabla^{\cal P}$ and $\Delta^{\cal P}$
as well as the scalar product $\SP{\cdot}{\cdot}$ like in the common
case of two-dimensional cartesian coordinates, these equations can also be rewritten as
\numparts \begin{eqnarray}\fl
\rho \left[ \cosh\Lambda \nabla^{\cal P} W  + \rme^{2W}\frac{\sinh\Lambda}{2R} \nabla^{\cal P} N
  - \frac{\cosh\Lambda-1}{2R} \nabla^{\cal P} R \right] = 0 \label{EQ:redeq.a}
\\\fl
\Delta^{\cal P} R = 0\label{EQ:redeq.b}
\\\fl
\frac{1}{R} \DIV^{\cal P} \left[ 2R \nabla^{\cal P} W \right] + \frac{\rme^{4W}}{R^2}\SP{\nabla^{\cal P} N}{\nabla^{\cal P} N}
= \rme^{2(V-W)}\rho \cosh\Lambda \label{EQ:redeq.c}
\\\fl
\DIV^{\cal P} \left[\frac{\rme^{2W}}{R} \nabla^{\cal P} N\right] + \frac{2\rme^{2W}}{R} \SP{\nabla^{\cal P} W}{\nabla^{\cal P} N}
= -\rme^{2(V-W)}\rho \sinh\Lambda \, .\label{EQ:redeq.d}
\end{eqnarray} \endnumparts
It should be emphasized, that similar equations could be derived for a fluid with pressure.
Furthermore, these equations would also be valid in the vacuum case - however,
\eref{EQ:redeq.a} would then be trivially fulfilled.

In the upcoming calculations of the interior solution, the fact that \eref{EQ:redeq.a}
is not trivial plays an important role.

\section{Solutions of the Einstein equations}

When writing $\rho = \Delta^{\cal P} U$ for some Potential $U$, the structure of
equations (\ref{EQ:redeq.a}-$d$) is invariant with respect to coordinate transformations
$(\xi,\eta) = \Phi(\mu,\nu)$ where $\xi$ and $\eta$ are harmonic functions
satisfying $\xi_\mu = \eta_\nu$ and $\xi_\nu = -\eta_\mu$.
In other words, coordinate transformations that can be written as
$\xi+i\eta = f(\mu+i\nu)$ for some holomorphic function $f$.

A helpful consequence of this fact is that $R$, together with a function $Q$ obeying
$Q_\mu = R_\nu$ and $Q_\nu = -R_\mu$, can also be considered as coordinates,
since by \eref{EQ:redeq.b} $R$ is harmonic in $\mu$ and $\nu$.
Vice versa, $R$ can be taken as any harmonic function in $\mu$ and $\nu$,
since a choice of $R$ is nothing else but a choice of the parametrization.
The only restriction is that $R$ should remain positive, when this function
is considered as a radial distance from the rotation axis, as it will be done in the following.
This interpretation of $R$ is quite natural, because in the case
$W=0$, $N=0$ the metric looks like the flat Minkowski metric in the coordinates
$(x,y,z) = (R(\mu,\nu)\cos\varphi, R(\mu,\nu)\sin\varphi, Q(\mu,\nu))$.
In fact, this is not a real restriction to the choice of $R$, but merely a restriction on
the range of the values of the coordinates $\mu$ and $\nu$.
By this interpretation of $R$, the associated function $Q$ can be interpreted
as the distance from the hypersurface defined by $Q=0$, that can be set as the symmetry plane -
if the solution is assumed to have a reflection symmetry.
This makes $R$ and $Q$ interpretable as some kind of cylindrical coordinates,
called Weyl's canonical coordinates.

\subsection{Rigid rotation and static solutions}
\label{SEC:rigi}

For $\Omega \equiv 0$, which due to \eref{EQ:X} and \eref{EQ:angovelo} is equivalent to $\Lambda \equiv 0$,
the dust is not rotating with respect to the background of fixed stars and can thus be
called static. 
This section is dedicated to the relation between solutions describing rigid rotation
and stationary solutions by the use of co-moving coordinates.
This transformation makes (\ref{EQ:redeq.a}-$d$) look less complicated
for $\Lambda = 0$.
This procedure to find solutions has also been used in \cite{2005astro.ph..7619C,
2006astro.ph..3791B,2006astro.ph..2519B}.
Furthermore, it will be shown that solutions describing rigid rotation obtained
from static solutions by this procedure will lead to unphysical metrics unless the
metric functions obey certain inequalities.

\subsubsection{About co-moving coordinates}
\label{SUBSEC:co-moving}

Solutions of (\ref{EQ:redeq.a}-$d$) describing rigid rotation are obtained by setting the 4-velocity
$Z$ parallel to $\partial_t - \Omega \partial_\phi$ where $\Omega$ is constant, but nonzero.
Since matter is assumed to move on timelike geodesics, condition \eref{EQ:rot_cond} must be
satisfied where $\rho \neq 0$.
However, for any asymptotically flat solution the right hand side of \eref{EQ:rot_cond}
should go towards $1$ for large values of $R$ or $Q$, whereas the left hand side can be
arbitrarily large.
This implies that the rotating dust cloud must be contained in a domain with $R<R_{max}$.
Note that the conclusions above also hold when pressure is included.

The co-moving coordinates are chosen in a way to have $\partial_{\wt{t}} = \partial_t - \Omega \partial_\varphi$.
The simplest way to accomplish this is to set $\wt{\varphi} := \varphi - \Omega t$ and to take
the identity mapping for the other coordinates.
Then also $\wt{Z} = \wt{E}_0$ and the transformed metric looks like
\begin{equation}\fl\label{EQ:metric_tilde}
\eqalign{g =&\left[ \rme^{2W}(1+\Omega N)^2 - \rme^{-2W}
\Omega^2 R^2 \right] \rmd \wt{t}^2 - 2\left[ \rme^{2W}(1+\Omega N)N - \rme^{-2W}
\Omega R^2 \right] \rmd \wt{t}
\, \rmd \wt{\varphi}  \\
&\quad - (\rme^{-2W}R^2 - \rme^{2W}N^2) \rmd \wt{\varphi}^2 - \rme^{2(V-W)}(\rmd \wt{\mu}^2 -
N \rmd \wt{\nu}^2)\, .}
\end{equation}
Thus, where \eref{EQ:rot_cond} holds, $\partial_{\wt{t}}$ is timelike and the transformed metric
\eref{EQ:metric_tilde} can be written in the same form as \eref{EQ:metric} with the metric functions
\numparts \begin{eqnarray}
\rme^{2\wt{W}} &= \rme^{2W}(1+\Omega N)^2 - \rme^{-2W} \Omega^2 R^2 \label{EQ:metric_funcs_tilde.a} \\
\wt{N}&= \frac{\rme^{2W}(1+\Omega N)N - \rme^{-2W} \Omega R^2}{
  \rme^{2W}(1+\Omega N)^2-\rme^{-2W} \Omega^2 R^2}\label{EQ:metric_funcs_tilde.b} \\
\rme^{2\wt{V}} &= \frac{\rme^{2V}}{\rme^{2W}(1+\Omega N)^2 - \rme^{-2W} \Omega^2 R^2} \label{EQ:metric_funcs_tilde.c} \\
\wt{R}^2 &= R^2  \, .  \label{EQ:metric_funcs_tilde.d}
\end{eqnarray} \endnumparts
Hence, by the use of co-moving coordinates all solutions describing rigidly rotating dust
can be reduced to the stationary case $\Lambda = 0$.
The transformation back into non co-moving coordinates can be obtained from the above by simply
replacing $\Omega$ with $-\Omega$ as
\numparts \begin{eqnarray}
\rme^{2W}  &= \rme^{2\wt{W}}(1-\Omega \wt{N})^2 - \rme^{-2\wt{W}} \Omega^2 \wt{R}^2 \label{EQ:metric_funcs_back.a}  \\
N &= \frac{\rme^{2\wt{W}}(1-\Omega \wt{N})\wt{N} + \rme^{-2\wt{W}} \Omega \wt{R}^2}{
  \rme^{2\wt{W}}(1-\Omega \wt{N})^2 - \rme^{-2\wt{W}} \Omega^2 \wt{R}^2}\label{EQ:metric_funcs_back.b}  \\
\rme^{2V}  &= \frac{\rme^{2\wt{V}}}{\rme^{2\wt{W}}(1-\Omega \wt{N})^2
  - \rme^{-2\wt{W}} \Omega^2 \wt{R}^2} \label{EQ:metric_funcs_back.c}  \\
R^2  &= \wt{R}^2 \, .\label{EQ:metric_funcs_back.d}
\end{eqnarray} \endnumparts
It can be read directly from \eref{EQ:metric_funcs_tilde.a} that $\rme^{2\wt{W}} > 0$ is equivalent
to \eref{EQ:rot_cond} and similarly, according to \eref{EQ:metric_funcs_back.a}, $\rme^{2W}>0$ is
equivalent to \eref{EQ:rot_cond} for the metric functions (\ref{EQ:metric_funcs_tilde.a}-$d$) with
$-\Omega$ instead of $\Omega$. Hence, a stationary solution satisfying
\numparts \begin{eqnarray}
0 &< \rme^{2\wt{W}}  \label{EQ:rigi_cond.a}  \\
0 &< \rme^{2\wt{W}}(1-\Omega \wt{N})^2 - \Omega^2 R^2 \rme^{-2\wt{W}}\label{EQ:rigi_cond.b}
\end{eqnarray} \endnumparts
can be transformed into an interior solution describing rigidly rotating matter with
timelike $\partial_t$ and $Z$.
Of course, by what was said at the beginning of this paragraph, such a solution would
not be asymptotically flat or make no sense as a global solution from a physical
point of view. It was already shown in \cite{2006astro.ph..3791B}
that globally defined rigid rotation solutions are plagued by unphysical properties.
Further, it might also be convenient to demand that $\Omega N$ and $\Omega \wt{N}$ should be
of positive sign in order to prevent the solution from containing some counter-rotating
sources of matter or singularities. This leads to the additional conditions
\numparts \begin{eqnarray}
0 \leq \Omega \wt{N} \label{EQ:rigi_cond_plus.a} \\
\rme^{2\wt{W}} (1-\Omega \wt{N})^2 - \rme^{-2\wt{W}} \Omega^2 R^2
  \leq \rme^{2\wt{W}}(1-\Omega \wt{N})  \, . \label{EQ:rigi_cond_plus.b}
\end{eqnarray} \endnumparts
A further condition for a reasonable solution would also be to demand that $\partial_\varphi$
should be spacelike, which is equivalent to demand $\partial_{\wt{\varphi}}$ being spacelike, i.e.
\begin{equation} \label{EQ:rigi_cond_phi}
\rme^{-2\wt{W}}\wt{R}^2 - \rme^{2\wt{W}}\wt{N}^2 \geq 0\, .
\end{equation}

\subsubsection{Einstein's equations for static dust}
\label{SUBSEC:staty}

In the static case, and for $\rho \neq 0$, \eref{EQ:redeq.a} requires setting
$W$ equal to a constant.
Since the addition of an arbitrary constant to $W$ would just result in a scale transformation,
this constant can be set equal to zero without any loss of generality.
Thus, it can be assumed $W \equiv 0$ henceforth.
With these simplifications (\ref{EQ:V.a},$b$) can be written as
\numparts \begin{eqnarray}\fl
V_\mu &= \frac{1}{2}\partial_\mu \ln (R_\mu^2+R_\nu^2)
  + \frac{ R_\mu N_\nu^2-R_\mu N_\mu^2-2R_\nu N_\mu N_\nu }
{4R(R_\mu^2+R_\nu^2)} \label{EQ:V_norot.a}
\\\fl
V_\nu &= \frac{1}{2}\partial_\nu \ln (R_\mu^2+R_\nu^2)
  + \frac{ -R_\nu N_\nu^2+R_\nu N_\mu^2-2R_\mu N_\mu N_\nu }
{4R(R_\mu^2+R_\nu^2)} \, . \label{EQ:V_norot.b}
\end{eqnarray} \endnumparts
Furthermore, (\ref{EQ:redeq.c},$d$) reduce to
\numparts \begin{eqnarray}
&\abs{\frac{\nabla^{\cal P}N}{R}}^2 = \rme^{2V}\rho
\label{EQ:Einstein_norot.a}\\
&\DIV^{\cal P} \left[\frac{\nabla^{\cal P} N}{R} \right] = 0 \label{EQ:Einstein_norot.b}
 \, .
\end{eqnarray} \endnumparts
Obviously, these equations are completely determined by the function $N$, which itself must
obey \eref{EQ:Einstein_norot.b}.
An important property of these equations is that, according to \eref{EQ:Einstein_norot.a},
the energy density $\rho$ is always positive. In \eref{EQ:redeq.c} this cannot be assumed
a priori - and as calculations will show, in the case of non-rigid rotation the mass
density can indeed become negative.

Some examples, how explicit solutions of that equation can found in some specific
parameterizations are given in appendix \ref{APP:sols}.
Another way to construct solutions of \eref{EQ:Einstein_norot.b} is mentioned in
\ref{SUBSEC:rigi_galaxy}.

\subsubsection{Rigidly rotating dust as a galaxy model}
\label{SUBSEC:rigi_galaxy}

This section about rigidly rotating dust solutions will be concluded by a short discussion about
the applicability of those metrics as galaxy models, as it has been done in
\cite{2005astro.ph..7619C,2006astro.ph..2519B}.
Since in both papers an attempt is made to give global solutions,
they surely must, as mentioned before, contain domains where the
metric becomes unphysical.
Where the problem lies in \cite{2005astro.ph..7619C,2006astro.ph..2519B}
will briefly be analyzed by the use of (\ref{EQ:rigi_cond.a},$b$) and (\ref{EQ:rigi_cond_plus.a},$b$).
In \cite{2005astro.ph..7619C,2006astro.ph..2519B} it has been set $\wt{W} = 0$.
Hence, \eref{EQ:rigi_cond.a} holds, so according to \eref{EQ:rigi_cond.b}
and (\ref{EQ:rigi_cond_plus.a},$b$),
$\Omega\wt{N}$ must be bounded by $0 \leq \Omega \wt{N} \leq 1$.
Thus, \eref{EQ:rigi_cond.b} will not hold if $R>R_{max}$ for some $R_{max}$.
This will happen for any choice of $\Omega \neq 0$.
As a result, the metric in non co-moving coordinates has a time-coordinate
which becomes spacelike as soon as $R>R_{max}$.

To avoid this problem, the metric for rigidly rotating dust must be restricted to a domain
with $R<R_{max}$ where (\ref{EQ:rigi_cond.a},$b$), (\ref{EQ:rigi_cond_plus.a},$b$)
and \eref{EQ:rigi_cond_phi} hold, and continued by another metric.
Remembering that \eref{EQ:rigi_cond_plus.a} is only necessary for rigidly rotating solutions,
the continued metric could also be a metric describing rotating dust,
but with non-rigid rotation and an angular velocity that falls off to zero for large values of $R$ and $Q$.
However, any transition into such an axisymmetric dust solution would still contain
another source of unphysical behavior, namely that circular orbits are still geodesics
according to \eref{EQ:divfree.b}.
This is only physically reasonable in the vicinity of the galactic plane \cite{2006astro.ph..1191C}.
Based on this observation, a further restriction of the solution to a domain with
$\abs{Q}<Q_{max}$, when $Q$ is interpreted as the distance from the galactic plane,
and a transition into a vacuum state appears to be the simplest choice to obtain physically
reasonable solutions.
Unfortunately, as will be shown later, this attempt will result in some
other singularities.

It has already been shown by Van Stockum \cite{VanStockum:1937prs} that interior solutions
for rigidly rotating dust can be obtained by axisymmetric solutions of Laplace's equation in
three dimensions. This relation does not only allow more general methods to find solutions of
\eref{EQ:Einstein_norot.b} than shown in \ref{APP:sols}, but can also give some deeper insight
into the behavior of the solution due to corresponding theorems about harmonic functions.
The relation mentioned above can be constructed as follows.
Using the linear mapping $J:\R^2\to\R^2$ describing a rotation of $\frac{\pi}{2}$, i.e.
$J(x,y) = (-y,x)$, an immediate consequence of \eref{EQ:Einstein_norot.b} is that,
according to Poincar\'{e}'s lemma, a function $H$ can locally be defined by
\begin{equation}  \label{EQ:H}
\nabla^{\cal P} H = \frac{1}{R} J\nabla^{\cal P} N\, ,
\end{equation}
up to an arbitrary constant.
Especially when $N$ is defined for $\mu,\nu \in U$ where $U$ is a simply connected domain,
$H$ can also be defined on whole $U$.
This function $H$ must then satisfy
\begin{equation}  \label{EQ:H_DEQ}
\DIV^{\cal P} \left[ R\nabla^{\cal P} H \right] = 0\, ,
\end{equation}
which is nothing else but Laplace's equation for an axisymmetric function in three dimensions
in the coordinates $(x,y,z) = (R(\mu,\nu)\cos\varphi, R(\mu,\nu)\sin\varphi, Q(\mu,\nu))$.
Two quite fundamental consequences of the relation \eref{EQ:H} are listed below.

One of the most important observations is a direct consequence of the fact that
harmonic functions have no isolated extrema.
Thus, any non-trivial globally defined asymptotically flat solution $\wt{N}$ of
\eref{EQ:Einstein_norot.b} must contain a singularity - if $\wt{N}$ is continuous,
then its first or second derivatives must contain some incontinuity.
This is not really astonishing, since such a solution would in fact describe a non-rotating
dust cloud without any inner pressure to keep the particles in their place.
Hence, from a physical point of view, the logical consequence is that there must be some kind
of singularity with a negative mass, as shown in \cite{2006astro.ph..3791B},
to prevent the cloud from collapsing. Furthermore, the presented discussion illustrates
that it would be a hopeless undertaking to model a spheroidal galaxy of type E0
in the Hubble sequence as gravitationally stabilized dust. 

Another consequence is that the whole spacetime cannot be covered singularity-free by
restricting the rotating dust solution to an axisymmetric region 
where $\mu$ and $\nu$ live on a domain on which Poincar\'{e}'s lemma can be applied.
Due to \eref{EQ:Einstein_norot.a}, such a transition
would yield $N=const.$ and thus $H=const.$ on the border of the domain containing the
dust cloud.
As a consequence, $H$, and therefore $N$, must be constant on the whole domain,
implying that there is also $\rho=0$, i.e. a vacuum, inside the domain according to
\eref{EQ:Einstein_norot.a}.
For the use as galaxy models, it would probably be sufficient to just consider interior
solutions on simply connected domains.
Nevertheless, the nonexistence of transitions into a vacuum for more general measurable domains
can also be verified under some further assumptions as follows.
Let $\mu,\nu \in U$, where $U$ is bounded with a piecewise smooth Lipschitz boundary $\partial U= A_U + \Gamma$, where $A_U$ is the intersection of $U$ with
the rotation axis and $\Gamma$ the 'real' boundary of the domain containing the rigidly rotating
dust cloud. Furthermore, $N\slash R$ and $\nabla^{\cal P}N\slash R$ should be well-defined as Sobolev
$\mbox{H}^1$ functions on $U$ and $\SP{\nabla^{\cal P}N}{n} \equiv 0$ on $A_U$, where $n$ is the unit
normal direction. These assumptions are not too devious, since they guarantee that the metric does
not become singular on the rotation axis (if $W$ satisfies similar conditions) and also, if the
ratio $N\slash R$ is small enough, that $\partial_\varphi$ remains spacelike.
By using \eref{EQ:Einstein_norot.b} and Gauss' theorem, it follows
\begin{equation}\fl \label{EQ:N_int}
\eqalign{\INT{U}{}{\mu}\rmd\nu \frac{\SP{\nabla^{\cal P}N}{\nabla^{\cal P}N}}{R}
&= \INT{U}{}{\mu}\rmd\nu N\DIV^{\cal P} \left[ \frac{\nabla^{\cal P}N}{R} \right] \\
& \qquad + \INT{U}{}{\mu}\rmd\nu \frac{\SP{\nabla^{\cal P}N}{\nabla^{\cal P}N}}{R} \\
&= \INT{U}{}{\mu}\rmd\nu \DIV^{\cal P} \left[ N \frac{\nabla^{\cal P}N}{R} \right] \\
&= \INT{\partial_U}{}{s(\mu,\nu)} \frac{N}{R} \SP{\nabla^{\cal P}N}{n} \\
&= \INT{\Gamma}{}{s(\mu,\nu)} \frac{N}{R} \SP{\nabla^{\cal P}N}{n} \, .}
\end{equation}
A direct consequence of demanding $\nabla^{\cal P}N \equiv 0$ on $\Gamma$ is
$\nabla^{\cal P}N \equiv 0$ and therefore a vacuum inside $U$.

As the calculations above show, a cloud of rigidly rotating dust on a bounded domain cannot exist without any forces stabilizing that matter configuration. Furthermore,
some embedded vacuum solutions, even including exotic objects like toroidally shaped black holes,
would not do the trick, since any continuation into a vacuum solution would be insufficient.
Thus, only continuations into other nonempty solutions, like e.g. a dark matter halo,
are possible choices.
Hence, claiming that solutions considered in this section are realistic models for the galactic disk
could rather be viewed as an argument in favor of the existence of dark matter than against it.

\subsection{Solutions describing arbitrary rotation}
\label{SEC:arbirot}

In the following it is presupposed that $\Lambda\neq 0$ in all points.
The corresponding interior solutions for non-rigidly rotating dust have already been
elaborated by Winicour in 1975 \cite{1975JMP....16.1806W}, a derivation adapted to
the form of the metric used in this paper is given in \ref{APP:arbirot}.
The solutions are completely characterized by an arbitrary positive function of one
variable $q$ and a solution of \eref{EQ:Einstein_norot.b} $\Psi$.
Furthermore, $W$ and $N$ can be expressed as functions of $R$ and $X$ \eref{EQ:X} as
\begin{eqnarray}
\rme^{-2W} = \frac{q(X)}{1-R^2X^2}\label{EQ:W1} \\
N = q(X) \frac{R^2X}{1-R^2X^2} + \frac{1}{2} \INT{X_0}{X}{y} \frac{q'(y)}{y} + N_0 \, . \label{EQ:N}
\end{eqnarray}
With constants $X_0$ and $N_0$.
The value of $X$ is related to $q$ and $\Psi$ by
\begin{equation} \label{EQ:Psi}
\Psi = R^2 f(X) + h(X)  \, ,
\end{equation}
where
\begin{equation}\label{EQ:f}
f'(X) = 2-X^2 h'(X) \, , \quad h'(X) = \frac{q'(X)}{Xq(X)}  \, .
\end{equation}
Since $1$ and $R^2$ are solutions of \eref{EQ:Einstein_norot.b}, it is not a real obstacle
that, according to \eref{EQ:f}, $f$ and $h$ are only defined up to some constants.
This could e.g. be fixed by demanding $f(X_0)=0$ and $h(X_0)=0$.
The important question is now, whether \eref{EQ:Psi} can be solved for $X$.
Assuming
\begin{equation}  \label{EQ:solve_cond}
R_0^2 f'(X_0) + h'(X_0) = 2R_0^2 + (1-R_0^2X_0^2)h'(X_0) \neq 0
\end{equation}
for some value $R_0$, the theorem of implicit functions guarantees that \eref{EQ:Psi} can really
be solved for $X$ as long as $\Psi$ and $\abs{R-R_0}$ are small enough, since for
$X = X_0$ the solution of \eref{EQ:Psi} is given by $\Psi \equiv 0$ .
Based on the theorem, the gradient $\nabla^{\cal P} X$ can be expressed as
\begin{equation}  \label{EQ:grad_X}
\nabla^{\cal P} X = \frac{1}{2\chi(R,X)}\left[ \nabla^{\cal P} \Psi - 2f(X) R \nabla^{\cal P} R \right] \, ,
\end{equation}
with $\chi$ as defined in \eref{EQ:chi}.
Because the solution $\Psi \equiv 0$ for $X = X_0$ is given for any choice for $R$,
\eref{EQ:Psi} can be in general only be solved for all values of $R$ if $\Psi$ satisfies
some condition $C(R) > \abs{\Psi}$ for a positive function $C$ depending on $q$.
Nevertheless, in the special case where $h'$ is positive with $h'(X_0)>0$ and $X^2h'(X)<\delta<2$,
condition \eref{EQ:solve_cond} would always be true and \eref{EQ:Psi} can be solved
globally without any further restrictions on $\Psi$.

Furthermore, the energy density can be expressed as
\begin{equation}\fl\label{EQ:rho_X}
\eqalign{\rme^{2(V-W)}\rho
&=  \frac{q'(X)^2 - R^4 \left[2Xq(X)-X^2q'(X)\right]^2}{4q^2R^2X^2} \abs{\nabla^{\cal P} X}^2\\
&=  \frac{h'(X)^2 - R^4 f'(X)^2 }{4R^2} \abs{\nabla^{\cal P} X}^2  \, .}
\end{equation}
By means of \eref{EQ:grad_X}, $\rho$ can also be written as
\begin{equation}\fl\label{EQ:rho}
\eqalign{\rme^{2(V-W)}\rho
&=  \frac{(1+R^2X^2)q'(X) - 2R^2X q(X)}{(1-R^2X^2)q'(X) + 2R^2X q(X)}\abs{ \frac{\nabla^{\cal P} \Psi}{2R} - f(X) \nabla^{\cal P} R  }^2  \\
&=  \frac{(1+R^2X^2)h'(X) - 2R^2}{(1-R^2X^2)h'(X) + 2R^2}\abs{ \frac{\nabla^{\cal P} \Psi}{2R} - f(X) \nabla^{\cal P} R  }^2 \, .}
\end{equation}
Since $X$ and $R$ only depend on the variables $\mu$ and $\nu$, $\rho$ can also easily be
expressed by using the gradient assigned to the metric $g$ instead of $\nabla^{\cal P}$ as
\begin{equation}\fl  \label{EQ:rho_GR}
\eqalign{\rho&=  -\frac{h'(X)^2 - R^4 f'(X)^2 }{4R^2} g(\mbox{grad}\, X, \mbox{grad}\, X) \\
 &=  -\frac{1}{4R^2} g\left( \mbox{grad}\, h(X) + R^2 \mbox{grad}\, f(X), \mbox{grad}\, h(X) - R^2 \mbox{grad}\, f(X) \right) \, .}
\end{equation}

\section{Discussion}

The function $q$ has been introduced as a degree of freedom. 
Since $q$ depends on the angular momentum $X$, and $W$ is constant in the stationary case $X=0$,
the function $q$ must contain some information about the rotation law of the dust cloud.
This can be seen in a more formal way as follows.
According to \eref{EQ:X_angovelo} as well as \eref{EQ:W1} and \eref{EQ:N}, the angular
velocity $\Omega$ obeys the relation
\begin{equation}  \label{EQ:q_angovelo}
\frac{1}{\Omega} = \frac{q(X)}{X} - \frac{1}{2} \INT{X_0}{X}{y} \frac{q'(y)}{y} - N_0 \, .
\end{equation}
Taking the gradient of this equation yields
\begin{equation}\fl  \label{EQ:q_grad_angovelo}
\mbox{grad}\, \frac{1}{\Omega}
  = \left[ -\frac{q(X)}{X^2} + \frac{q'(X)}{2X} \right] \mbox{grad}\,X
  = \frac{1}{2} \left[ 2q(X) - X q'(X) \right] \mbox{grad}\,\frac{1}{X}  \, .
\end{equation}
Thus, the value $2q(X) - X q'(X)$ can be taken as an indicator how strongly the rotation
of the matter deviates from rigid rotation.
Especially if that factor vanishes the rotation is rigid - of course, there would also be no
non-rigidly rotating matter if $\mbox{grad}\, X = 0$, but because of \eref{EQ:rho_X} there
would be no matter at all.

As aforementioned in \ref{SEC:rigi}, the energy density can be negative if $W$ is not constant.
According to \eref{EQ:rho_X}, the energy density is positive only as long as
\begin{equation}  \label{EQ:rho_pos1}
h'(X)^2 - R^4 f'(X)^2 \geq 0  \, .
\end{equation}
This relation can also be written as
\begin{equation}  \label{EQ:rho_pos2}
R^2 \leq \abs{ \frac{h'(X)}{2-X^2h'(X)} } = \abs{ \frac{q'(X)}{X[2q-Xq'(X)]} }  \, ,
\end{equation}
provided the denominator is different from zero.
The cases where \eref{EQ:rho_pos2} is not well defined are $X=0$ and $2q-Xq'(X)=0$,
i.e. when there is no or just rigid rotation -
presupposing $q'$ as continuous.
Hence, if the rotation is differential, the deviation from rigid rotation implies
a restriction on the maximal radius of the rotating dust cloud, which is inversely
proportional to the deviation.
This inequality above is a further restriction for $R$ besides $\abs{XR} < 1$,
since for arbitrary $h'$ the latter cannot be derived from \eref{EQ:rho_pos1} or \eref{EQ:rho_pos2}.
Especially, when $\frac{1}{X^2} > R^2 > \abs{h'(X)}$, $\rho$ will be negative.
Hence, all physically reasonable solutions must obey $R^2 \leq \abs{h'(X)}$.

However, when assuming $0\leq X^2 h'(X) < 1$ then \eref{EQ:rho_pos2} also implies
$\abs{RX} < 1$. This implication is strict in a sense that for any $\wt{X}$ with
$\wt{X}^2 h'(\wt{X}) \geq 1$ there exists an $\wt{R}$ that obeys
\eref{EQ:rho_pos2} but $\abs{\wt{R}\wt{X}} \geq 1$.
A direct consequence of the discussion above is that all matter must be contained in
a domain with a finite radius when $h'$ is bounded, as in the case of rigid rotation.

The restriction of the dust cloud in the radial $R$ direction will also follow if the
solution converges smoothly enough to the asymptotic flat space.
Since for $R\to\infty$ it is necessary to have $X\slash\Omega \to 1$, implying by
\eref{EQ:q_angovelo} that $q \to q_{\infty}$ with $q_{\infty}(X) = 1 - (N_0\slash X_0)X^2$
for $X_0 \neq 0$ and $q_{\infty}\equiv 1$ otherwise, because $X_0=0$ implies $N_0=0$
when asymptotically $X \to 0$.
According to \eref{EQ:rho_pos2} $R^2$ must then be bounded by $\abs{N_0\slash X_0}$
or $0$ respectively, which is a contradiction to $R\to\infty$.
Thus, the matter cannot be contained in a region with arbitrarily large values for $R$.

It was mentioned that
\begin{equation}  \label{EQ:rho_zero}
(1+R^2X^2)h'(X) - 2R^2 = h'(X) - [2 - X^2h'(X)]R^2 = 0
\end{equation}
can have nontrivial solutions.
These mark the points of the spacetime where $\rho$ is vanishing, assuming
$\frac{1}{R}\nabla^{\cal P}\Psi$ is not singular in these points.
Furthermore, according to \eref{EQ:rho}, the equation
\begin{equation}  \label{EQ:rho_sing}
(1-R^2X^2)h'(X) + 2R^2 = h'(X) + [2 - X^2h'(X)]R^2 = 0
\end{equation}
marks the points where the energy density is potentially singular.
Regarding \eref{EQ:solve_cond}, then \eref{EQ:rho_sing} marks the values $(R,X)$
where the implicit function theorem cannot be applied. This is because
$\nabla^{\cal P} X$ would become singular and additionally the implicit equation \eref{EQ:Psi}
might possibly have a branch-point there.

If $R > 0$, the energy density cannot be positive for $h'(X)=0$ for obvious reasons.
Thus it can be assumed that $h'$ will not change the sign on the whole domain with $\rho>0$.
To avoid singularities and being able to find coordinates, where $\rho$ can vanish to zero,
it is (according to \eref{EQ:rho_zero} and \eref{EQ:rho_sing}) quite advantageous to choose
$h'$ as positive on that domain.
Remembering that $R$ and $X$ are both functions of the two variables $\mu$ and $\nu$, and that
\eref{EQ:redeq.a} is only non-trivial for $\rho \neq 0$, \eref{EQ:rho_zero} can be seen as the specification for the border of the region where the interior
solution makes physically sense.
By choosing $q$, or $h'$ respectively, and $\Psi$ in a suitable way, interior metrics can
be found, where $\rho$ indeed vanishes on the surface of a bounded domain containing the dust,
which could be seen as the edge of the galactic disk.
This situation allows a somewhat more formalistic approach to the dark matter problem in terms
of general relativity, namely whether a smooth transition from the rotating dust into an
asymptotically flat vacuum solution can be found or does not exist.
Though some axisymmetric vacuum solutions are known \cite{2003esef.book.....S},
especially including solutions outside an infinitesimal disk
\cite{1998rdgm.conf.....M,2000GReGr..32.1365A,1999PhRvL..83.2884K},
which could yield good approximations
for the vacuum solution outside a galactic disc (should it exist) there does not yet seem
to be a theorem that could guarantee either the possibility or the impossibility of a
smooth transition.
However, what has been exhibited by recent developments
\cite{1998CMaPh.190..509S,2002PhRvD..65d4006A,1998MPLA...13.1509M} is that the asymptotically
flat vacuum solution outside a bounded region for given Dirichlet boundary data is unique
and the existence of a solution only seems to be guaranteed, if these data satisfy certain
conditions. Furthermore, in \cite{2002PhRvD..65d4006A} an
approximation scheme was presented, how solutions can be obtained from
it's Dirichlet boundary data.
Since smooth transitions not only require a correspondence of the Dirichlet data at the boundary,
but also of the Neumann boundary data, the problem of finding an attached vacuum solution
for an arbitrary choice of $q$ and $\Psi$ is overdetermined and would thus in general create
some kind of mass shell at the boundary, as it also was mentioned in \cite{1998MPLA...13.1509M}.

\subsection{The limit of slow rotation}
\label{SUBSEC:slowrot}

In the case of slow rotation where $X$ is small, the solution in \ref{SEC:arbirot}
should in some way be well approximated by a static solution, as described in \ref{SUBSEC:staty}.
Thus, a solution with an arbitrary rotation law will be expanded in $X$ to lowest order.

In the following it is assumed that $X_0=0$ and $q(0) = 1$.
In fact, $q(0)$ can be taken as any nonzero number, since multiplication of $q$ with
some nonzero number would simply result in a scale transformation.
The expansion of \eref{EQ:W1} and \eref{EQ:N} then yields
\begin{equation}\label{EQ:W_exp}
\rme^{2W} = 1 + \Or(X^2)
\end{equation}
and
\begin{equation}\label{EQ:N_exp}
N = \frac{\Psi}{2} + N_0 + \Or(X^2) \, ,
\end{equation}
where \eref{EQ:Psi} has been used.

Furthermore, an expansion of \eref{EQ:rho} will unveil the result
\begin{equation}  \label{EQ:rho_exp}
\rme^{2V}\rho = \frac{h'(0) - 2R^2}{h'(0) + 2R^2}\abs{ \frac{\nabla^{\cal P} \Psi}{2R} }^2 + \Or(X)\, .
\end{equation}
Thus, when the rotation is rather slow and $R$ is small in comparison to $h'(0)$ the metric
will indeed decay to a static solution as obtained in \ref{SUBSEC:staty}.
Even further, \eref{EQ:rho_exp} also shows that $\sqrt{\frac{h'(0)}{2}}$ can
be taken an indicator for the radius of the dust cloud.

\subsection{Back to rigid rotation}

Though the method of using co-moving coordinates described in \ref{SUBSEC:co-moving}
might be the easier attempt to find solutions describing rigid rotation, those solutions
must also be included in those from \ref{SEC:arbirot} - in non co-moving coordinates.
According to \eref{EQ:q_grad_angovelo}, rotation is rigid
for $q(X) = \frac{X^2}{\Omega^2}$, where an arbitrary factor has again been omitted.

By fixing the factors $X_0$ and $N_0$ according to \eref{EQ:q_angovelo},
it is now straightforward to calculate
\begin{eqnarray}
\rme^{2W} &= \frac{\Omega^2}{X^2} - \Omega^2R^2 \label{EQ:W_rigi}\\
N &= \frac{X}{\Omega^2(1-R^2X^2)} - \frac{1}{\Omega} \label{EQ:N_rigi}  \, .
\end{eqnarray}
Furthermore, \eref{EQ:Psi} takes the form
\begin{equation}\label{EQ:Psi_rigi}
\Psi = -2\frac{1}{X}
\end{equation}
and $\rho$ is given by
\begin{equation}  \label{EQ:rho_rigi}
\rme^{2(V-W)}\rho = \abs{ \frac{\nabla^{\cal P} \Psi}{2R} }^2\, .
\end{equation}
When comparing these results to the transformation rules
(\ref{EQ:metric_funcs_back.a}-$d$) with $\wt{W}=0$ and $\frac{1}{X} = \frac{1}{\Omega}(1-\Omega N)$,
the metric is indeed the solution for rigid rotation described in \ref{SEC:rigi}.

\section{Conclusions}

Besides the restriction that physically realistic solutions describing axisymmetric
rotating dust should be restricted in the $Q$ direction because, as already
stated in \cite{2006astro.ph..1191C},
the assumption of circular flow is unrealistic far off the galactic plane, the
calculations have shown that it is also required to restrict the solution in
the $R$ direction.
This unmasks the solutions in \cite{2005astro.ph..7619C,2006astro.ph..2519B}
once again as unphysical, though they seem to work quite well for the interior of the
galactic disk.
However, according to the expansions in \ref{SUBSEC:slowrot}, the latter is not accidental,
but simply because those metrics approximate non-rigid rotation solutions to first order.

Furthermore, solutions describing rigid rotation cannot simply be matched with exterior
vacuum solutions without becoming singular, making them a quite useless mean to claim
the non-existence of dark matter, besides the fact that rigid rotation might
possibly be quite inappropriate to describe the galactic rotation law, even though
the measured rotation curves look like differential rotation due to the relativistic dragging
effect observed by rotating objects.

Even when non-rigid rotation of matter is allowed as in \ref{SEC:arbirot},
things do not really improve in an optimal way.
Though, in comparison to the case of rigid rotation, the interior solution can now indeed have
an energy density $\rho$ that
vanishes on a compact surface, the resulting problem to find a matching vacuum solution
is over-determined. Thus, the resulting metric would in general contain an infinitesimal
mass shell enveloping the galactic disk.
Nevertheless, it cannot be excluded at the current time, that smooth transitions could exist.
Of course, also transitions into other solutions, especially such with more degrees of
freedom should be analyzed in the near future, e.g. a dust solution where the $4$-velocity
$Z$ is not restricted to be a linear combination of just $\partial_t$ and $\partial_\varphi$.
But still, already this assumption would require that metrics of a more general form than
\eref{EQ:metric} have to be analyzed, making the chance to obtain some reasonable results
that would finally conclude the debate about the existence of dark matter,
quite difficult if not impossible at the current moment.

Nevertheless, the results exhibit that galaxies should be treated as much more complex
objects than just a circularly rotating cloud of dust and a final solution to match
different solutions of the Einstein equations would also give some definite answers
concerning the existence of dark matter. Though this answer is still outstanding,
the apparent incapability of solutions containing circularly rotating dust as the only source
of gravity to serve as global metrics might give a further indication that more physics is needed
to model galaxies, like pressure related phenomena, dark matter halos, effects caused by
a massive central black hole or even non-axisymmetric dynamics.

\appendix

\section{}
\label{APP:arbirot}

For $\rho \neq 0$, i.e. if matter is present, \eref{EQ:redeq.a} implies a non-trivial
relation between $\nabla^{\cal P} W$, $\nabla^{\cal P} N$ and $\nabla^{\cal P} R$ as
\begin{equation}  \label{EQ:div1}
\cosh\Lambda \nabla^{\cal P} W  + \rme^{2W}\frac{\sinh\Lambda}{2R} \nabla^{\cal P}
N  - \frac{\cosh\Lambda-1}{2R} \nabla^{\cal P} R = 0\, .
\end{equation}
Using the linear mapping $J:\R^2\to\R^2$ introduced in \ref{SUBSEC:rigi_galaxy},
a direct consequence of \eref{EQ:div1} is
\begin{equation}\fl  \label{EQ:div2}
\DIV^{\cal P} J\left[ \cosh\Lambda \nabla^{\cal P} W  + \rme^{2W}\frac{\sinh\Lambda}{2R}
\nabla^{\cal P} N - \frac{\cosh\Lambda-1}{2R} \nabla^{\cal P} R \right] = 0\, .
\end{equation}
This equation guarantees that \eref{EQ:div1} can be solved for one of the values
$W$, $N$ or $R$, i.e. the integrability condition for $\nabla^{\cal P} W$,
$\nabla^{\cal P} N$ or $\nabla^{\cal P} R$ holds.

Calculating the application of $\DIV^{\cal P}$ in \eref{EQ:div2} and using \eref{EQ:div1}
to get rid of the term $\nabla^{\cal P} N$ leads to
\begin{equation}  \label{EQ:W-eq}
\SP{ \frac{\nabla^{\cal P} R}{R} - \frac{\nabla^{\cal P} \Lambda}{\sinh\Lambda} }{
J\left[\nabla^{\cal P} W - \frac{\nabla^{\cal P} R}{2R} -
\frac{\cosh\Lambda \nabla^{\cal P} \Lambda}{2\sinh\Lambda}\right] } = 0  \, .
\end{equation}
This equation is solved, when $W$ is written as
\begin{equation}  \label{EQ:W}
W = -\frac{1}{2} \ln\left[ \frac{1+\cosh\Lambda}{2} q(X) \right]\, ,
\end{equation}
where $q:\R\to\R_{\geq 0}$ is an arbitrary differentiable function and $X$ is
the function defined in \eref{EQ:X}.
As it will turn out, it is convenient to use the function $X$ instead of
$\Lambda$ or $\Omega$.
Hence, the following relations are useful
\begin{eqnarray}
\sinh\Lambda = \frac{2RX}{1-R^2X^2}  \label{EQ:X_Lambda.a}\\
\cosh\Lambda = \frac{1+R^2X^2}{1-R^2X^2} \, . \label{EQ:X_Lambda.b}
\end{eqnarray}
Now \eref{EQ:div1}, \eref{EQ:W} and (\ref{EQ:X_Lambda.a},$6$) can be used to calculate
\begin{eqnarray}\fl \label{EQ:grad_W}
\nabla^{\cal P} W = \frac{1}{2R} \left[ \nabla^{\cal P} R - \frac{1+R^2X^2}{1-R^2X^2}
\nabla^{\cal P} R - \chi(R,X) \frac{2X}{1-R^2X^2} \nabla^{\cal P} X \right] \\
\fl \label{EQ:grad_N}
\nabla^{\cal P} N = \frac{q(X)}{1-R^2X^2} \left[ \frac{2RX}{1-R^2X^2} \nabla^{\cal P} R
 + \chi(R,X) \frac{1+R^2X^2}{1-R^2X^2} \nabla^{\cal P} X \right]\, ,
\end{eqnarray}
where $\chi(R,X)$ is defined by
\begin{equation}  \label{EQ:chi}
\chi(R,X) = \frac{ (1-R^2X^2)q'(X) + 2R^2X q(X) }{2Xq(X)}\, .
\end{equation}
From this also follow \eref{EQ:W1} and \eref{EQ:N}.

The calculations above show that for any differentiable function $q$ and a parametrization
via $R$, the solution is determined by the function $X$,
as the solution for rigid rotations was
completely determined by a solution of \eref{EQ:Einstein_norot.b}.
Due to (\ref{EQ:redeq.c},$d$), the function $X$ cannot be chosen arbitrarily.
The crucial point is now to derive an equation for $X$ from (\ref{EQ:redeq.c},$d$).
As calculations will show, $X$ is related to a solution of \eref{EQ:Einstein_norot.b}.

The procedure to derive the mentioned equation for $X$ is as follows.
First, from (\ref{EQ:redeq.c},$d$) a new equation can be obtained which does
not contain the term $\rho$ due to the the combination
$\eref{EQ:redeq.c}\sinh\Lambda+\eref{EQ:redeq.d}\cosh\Lambda=0$.
This reads as
\begin{equation}\fl \label{EQ:match1}
\eqalign{ 0 &=\sinh\Lambda \left( \frac{1}{R} \DIV^{\cal P} \left[ 2R \nabla^{\cal P} W \right]
 + \frac{\rme^{4W}}{R^2}\SP{\nabla^{\cal P} N}{\nabla^{\cal P} N} \right) \\
& \qquad + \cosh\Lambda \left( \DIV^{\cal P} \left[\frac{\rme^{2W}}{R} \nabla^{\cal P}
 N\right]+ \frac{2\rme^{2W}}{R} \SP{\nabla^{\cal P} W}{\nabla^{\cal P} N}\right)\\
&=\frac{\sinh\Lambda}{R} \DIV^{\cal P} \left[ 2R \nabla^{\cal P} W \right]
 + \frac{\cosh\Lambda}{R} \DIV^{\cal P} \left[\rme^{2W} \nabla^{\cal P} N \right]  \\
& \qquad + \sinh\Lambda \frac{\rme^{4W}}{R^2}\SP{\nabla^{\cal P} N}{\nabla^{\cal P} N}
 - \cosh\Lambda \frac{1}{R^2} \SP{\nabla^{\cal P} R}{\frac{\rme^{2W}}{R} \nabla^{\cal P} N}  \\
& \qquad + \cosh\Lambda \frac{2\rme^{2W}}{R} \SP{\nabla^{\cal P} W}{\nabla^{\cal P} N} \, .}
\end{equation}
When using \eref{EQ:div1} in order to reduce the number of terms, this can be rewritten as
\begin{equation}\fl
\eqalign{ 0 &=  \frac{\sinh\Lambda}{R} \DIV^{\cal P} \left[ 2R \nabla^{\cal P} W \right] +
\frac{\cosh\Lambda}{R} \DIV^{\cal P} \left[\rme^{2W} \nabla^{\cal P} N \right]  \\
&\qquad - \frac{\rme^{2W}}{R^2}\SP{\nabla^{\cal P} R}{\nabla^{\cal P} N} \, .}
\end{equation}
Now, the terms $\cosh\Lambda$ and $\sinh\Lambda$ can be replaced by \eref{EQ:X_Lambda.a} and
\eref{EQ:X_Lambda.b} yielding
\begin{equation}\fl  \label{EQ:step_1}
\eqalign{ 0 &=  \frac{2X}{R(1-R^2X^2)} \DIV^{\cal P} \left[ 2R \nabla^{\cal P} W \right]
  + \frac{1+R^2X^2}{R(1-R^2X^2)} \DIV^{\cal P} \left[\rme^{2W} \nabla^{\cal P} N \right] \\
&\qquad - \frac{\rme^{2W}}{R^2}\SP{\nabla^{\cal P} R}{\nabla^{\cal P} N} \, .}
\end{equation}
By inserting \eref{EQ:W1}, \eref{EQ:grad_W} and \eref{EQ:grad_N}, \eref{EQ:step_1}
can be written as
\begin{equation}\fl  \label{EQ:step_2}
\eqalign{ 0 &=  \frac{2X}{R(1-R^2X^2)} \DIV^{\cal P} \left[ \nabla^{\cal P} R - \frac{1+R^2X^2}{1-R^2X^2} \nabla^{\cal P} R
  - \chi(R,X) \frac{2X}{1-R^2X^2} \nabla^{\cal P} X \right]\\
&\qquad + \frac{1+R^2X^2}{R(1-R^2X^2)} \DIV^{\cal P} \left[ \frac{2RX}{1-R^2X^2} \nabla^{\cal P} R
  + \chi(R,X) \frac{1+R^2X^2}{1-R^2X^2} \nabla^{\cal P} X \right]\\
&\qquad - \frac{1}{R^2}\SP{\nabla^{\cal P} R}{\frac{2RX}{1-R^2X^2} \nabla^{\cal P} R
  + \chi(R,X) \frac{1+R^2X^2}{1-R^2X^2} \nabla^{\cal P} X} \, . }
\end{equation}
This equivalent to
\begin{equation}\fl \label{EQ:match_2}
\eqalign{ 0 &=  \frac{1}{R} \DIV^{\cal P} \left[ \chi(R,X) \nabla^{\cal P} X \right]
  + \frac{1}{R} \SP{\nabla^{\cal P} R}{ \frac{1+R^2X^2}{1-R^2X^2}
\nabla^{\cal P}\left[ \frac{2RX}{1-R^2X^2} \right]}  \\
&\qquad - \frac{1}{R}\SP{\nabla^{\cal P} R}{\frac{2RX}{1-R^2X^2} \nabla^{\cal P}\left[
\frac{1+R^2X^2}{1-R^2X^2} \right] }
  - \frac{1}{R^2}\SP{\nabla^{\cal P} R}{\frac{2RX}{1-R^2X^2}
\nabla^{\cal P} R} \\
&\qquad - \frac{1}{R^2}\SP{\nabla^{\cal P} R}{\chi(R,X) \frac{1+R^2X^2}{1-R^2X^2}
\nabla^{\cal P} X} \, .}
\end{equation}
By straightforward calculation, this can be reduced to
\begin{equation}
0 =  \DIV^{\cal P} \left[ \frac{\chi(R,X)}{R} \nabla^{\cal P} X \right]
 + \left[ 2-X\frac{q'(X)}{q(X)} \right] \SP{\nabla^{\cal P} R}{\nabla^{\cal P} X} \, .
\end{equation}
Since $R$ is harmonic \eref{EQ:redeq.b}, the equation above is equal to
\begin{equation}  \label{EQ:match3}
0 = \DIV^{\cal P} \left[ \frac{\chi(R,X)}{R} \nabla^{\cal P} X  + f(X) \nabla^{\cal P} R \right]\, ,
\end{equation}
where $f$ is defined by \eref{EQ:f}.
With the substitution \eref{EQ:Psi}, what by use of \eref{EQ:chi} can also
be written as
\begin{equation} \label{EQ:Psi_2}
\Psi = \frac{1}{2} \INT{X_0}{X}{s} \chi(R,s) \, ,
\end{equation}
when assuming $f(X_0)=0$ and $h(X_0)=0$, \eref{EQ:match3} indeed turns out to be
equivalent to
\begin{equation} \label{EQ:Psi_eq}
\DIV^{\cal P} \left[ \frac{\nabla^{\cal P} \Psi}{R} \right] = 0\, .
\end{equation}
This is the same equation as \eref{EQ:Einstein_norot.b}.

By means of \eref{EQ:grad_X}, \eref{EQ:grad_W} and \eref{EQ:grad_N} can also be expressed as
\begin{eqnarray}\fl \label{EQ:grad_W_new}
\nabla^{\cal P} W = \frac{X}{1-R^2X^2} \left[ [f(X)-X]R\nabla^{\cal P} R - \frac{\nabla^{\cal P} \Psi}{2} \right]\\
\fl \label{EQ:grad_N_new}
\eqalign{ \nabla^{\cal P} N
&=  -\frac{q(X)}{(1-R^2X^2)} \Biggl[ \left( 2RX - \frac{1+R^2X^2}{1-R^2X^2} R f(X) \right) \nabla^{\cal P} R \\
&\qquad + \frac{1+R^2X^2}{1-R^2X^2} \frac{\nabla^{\cal P} \Psi}{2} \Biggr] \, .}
\end{eqnarray}
The energy density can be obtained by inserting \eref{EQ:match1} into \eref{EQ:redeq.c}.

\section{}
\label{APP:sols}

The differential equation \eref{EQ:Einstein_norot.b} plays
a central role when dealing with circularly rotating dust.
Even though solutions can be obtained from axisymmetric harmonic functions in
three dimensions, it would be quite practical to have some additional
explicit expressions, how solutions can be found by separation of variables,
when assuming that $R$ is separable in $\mu$ and $\nu$.
The so obtained solutions correspond to the most common coordinate systems that are suitable
for axisymmetric functions.

In the following some examples are given for $R$, $Q$ and \eref{EQ:Einstein_norot.b}
in different coordinate systems.
The designation of the systems refer to the parametrization in the case
of flat space, i.e. when $N=0$ and $W=0$.

\paragraph{Cylindrical coordinates :}
\begin{equation}\label{EQ:param_cyl}
R = \nu \, , \quad  Q = \mu
\end{equation}
\begin{equation} \label{EQ:E_cyl}
\nu\,\Psi_{\mu\mu} + \nu\,\Psi_{\nu\nu} - \Psi_\nu = 0
\end{equation}
Separating this equation with constant $\kappa^2, \kappa \in \C$ leads to
\begin{equation}\fl \label{EQ:cyl_sol}
\Psi_\kappa (\mu,\nu) = \left[ A_\kappa \rme^{\kappa \mu} + B_\kappa \rme^{-\kappa \mu} \right]
\left[ C_\kappa \nu J_1(\kappa \nu) + D_\kappa \nu N_1(\kappa \nu) \right]  \, ,
\end{equation}
where $J_1$ and $N_1$ are the Bessel function of the first and second kind with order $1$.
When further assuming reflection symmetry, \eref{EQ:cyl_sol} can be simplified to
\begin{equation}\fl \label{EQ:cyl_sol_sym}
\wt{\Psi}_\kappa (\mu,\nu) = \cosh(\kappa \mu) \left[ C_\kappa \nu J_1(\kappa \nu) + D_\kappa \nu N_1(\kappa \nu) \right]  \, .
\end{equation}
Solutions of this kind have been used for the modelling of
galactic rotation curves.
In \cite{2005astro.ph..7619C} it was claimed that such solutions are globally valid,
producing different sources of unphysical behavior, mainly because rigidly
rotating dust is described which must result in
some unphysical behavior, as mentioned in \cite{2006astro.ph..3791B} and discussed in
\ref{SUBSEC:rigi_galaxy}.
Nevertheless, as shown in \ref{SUBSEC:slowrot}, the results of Cooperstock \& Tieu
remain applicable in the vicinity of the galactic disc.

Another way to produce solutions of \eref{EQ:E_cyl} in cylindrical coordinates is
in the form of the integral representation
\begin{equation}\label{EQ:itregral_cyl}
\Psi(\mu,\nu) = \INT{\Gamma}{}{z} C(z) \frac{\nu^2}{((\mu-z)^2+\nu^2)^{\frac{2}{3}}}
\end{equation}
with an arbitrary function of a complex variable $C$ and a curve $\Gamma$ in the complex plane.
E.g. in \cite{2006astro.ph..2519B}, $|\Gamma| = \R$ and an even function $C$
was used to cover the whole spacetime.

\paragraph{Paraboloidal coordinates :}
\begin{equation}\label{EQ:param_para}
R = \mu\nu  \, , \quad  Q = \frac{1}{2}(\mu^2-\nu^2)
\end{equation}
\begin{equation} \label{EQ:E_para}
\mu\nu\,\Psi_{\mu\mu} + \mu\nu\,\Psi_{\nu\nu} - \nu \Psi_\mu - \mu \Psi_\nu = 0
\end{equation}
Separation with the constant $\kappa^2, \kappa \in \C$ leads to the solution
\begin{equation}\fl \label{EQ:para_sol}
\Psi_\kappa (\mu,\nu) = \mu \nu \left[ A_\kappa J_1(\kappa \mu) +
B_\kappa N_1(\kappa \mu) \right]
  \left[ C_\kappa J_1(\rmi\kappa \nu) +
D_\kappa N_1(\rmi\kappa \nu) \right]  \, .
\end{equation}
Reflection symmetry can be obtained by symmetrization like
\begin{equation} \label{EQ:para_sol_sym}
\wt{\Psi}_\kappa (\mu,\nu) = \Psi_\kappa (\mu,\nu) + \Psi_\kappa (\nu,\mu)  \, .
\end{equation}

\paragraph{Spheroidal coordinates :}
\begin{equation}\fl\label{EQ:param_sph}
R = R_0 \rme^\mu \sin\nu\, , \quad  Q = R_0 \rme^\mu \cos\nu
\end{equation}
\begin{equation}\fl\label{EQ:E_sph}
\sin\nu\,N_{\mu\mu} + \sin\nu\,N_{\nu\nu} - \sin\nu N_\mu - \cos\nu N_\nu = 0
\end{equation}
Solutions of this equation with separation constant $\lambda(\lambda-1) \in \C,
\Re \lambda>-1$ can be written as
\begin{equation}\fl \label{EQ:sph_sol}
\Psi_\lambda (\mu,\nu) = \cases{ \left[ A_0 \rme^{\mu} + B_0 \right]\left[ C_0 \cos
\nu + D_0 \right]  & $\lambda = 0$ \, , \\
\sin^2\nu \left[ A_\lambda \rme^{(\lambda+1) \mu} + B_\lambda \rme^{-\lambda \mu}
\right] \ast &  \\
\quad \left[ C_\lambda P'_\lambda(\cos\nu) + D_\lambda Q'_\lambda(\cos\nu)
\right]& $\lambda\neq 0$\, ,}
\end{equation}
where $P_\lambda$ and $Q_\lambda$ are Legendre functions of first and second kind.
For arbitrary $\lambda$, $P_\lambda$ and $Q_\lambda$ contain a logarithmic singularity
at $-1$ and $1$
respectively. Exceptions are $\lambda=l \in \N$, where $P_l$ is a polynomial of degree $l$.
When further assuming reflection symmetry, \eref{EQ:sph_sol} reduces to
\begin{equation}\fl \label{EQ:sph_sol_sym}
\wt{\Psi}_\lambda (\mu,\nu) = \cases{A_0 \rme^{\mu} + B_0& $\lambda = 0$ \, , \\
 \sin^2\nu P'_{2l+1}(\cos\nu) [ A_{2l+1} \rme^{(2l+2)\mu}&  \\
 \quad + B_{2l+1} \rme^{-(2l+1) \mu} ] & $\lambda = 2l+1$ \, .}
\end{equation}

\paragraph{Prolate spheroidal coordinates :}
\begin{equation}\fl\label{EQ:param_PSC}
R = R_0 \sinh\mu \sin\nu \, , \quad  Q = R_0 \cosh\mu \cos\nu
\end{equation}
\begin{equation}\fl\label{EQ:E_PSC}
\sinh\mu \sin\nu\,N_{\mu\mu} + \sinh\mu \sin\nu\,N_{\nu\nu}
  - \cosh\mu \sin\nu N_\mu - \sinh\mu \cos\nu N_\nu = 0
\end{equation}
Separating again with $\lambda(\lambda-1) \in \C, \Re \lambda>-1$ leads to the solution
\begin{equation}\fl \label{EQ:PSC_sol}
\Psi_\lambda (\mu,\nu) = \cases{ \left[ A_0 \cosh\mu + B_0 \right]\left[ C_0
\cos \nu + D_0 \right]& $\lambda = 0$ \, , \\
\sin^2\nu \sinh^2\mu \left[ A_\lambda P'_\lambda(\cosh\mu)
 + B_\lambda Q'_\lambda(\cosh\mu) \right]\ast  &  \\
\quad \left[ C_\lambda P'_\lambda(\cos\nu) + D_\lambda Q'_\lambda(\cos\nu) \right]
& $\lambda\neq 0$ \, .}
\end{equation}
$P_\lambda$ and $Q_\lambda$ are again solutions of the Legendre equation.
Here, with reflection symmetry and integer $\lambda$ \eref{EQ:PSC_sol} looks like
\begin{equation}\fl \label{EQ:PSC_sol_sym}
\wt{\Psi}_\lambda (\mu,\nu) = \cases{A_0 \cosh\mu + B_0& $\lambda = 0$ \, , \\
 \sin^2\nu \sinh^2\mu P'_{2l+1}(\cos\nu) [ A_{2l+1} P'_{2l+1}(\cosh\mu)  &  \\
 \quad + B_{2l+1} Q'_{2l+1}(\cosh\mu) ] & $\lambda=2l+1$\, .}
\end{equation}

\paragraph{Oblate spheroidal coordinates :}
\begin{equation}\fl\label{EQ:param_OSC}
R = R_0 \cosh\mu \cos\nu \, , \quad  Q = R_0 \sinh\mu \sin\nu
\end{equation}
\begin{equation}\fl\label{EQ:E_OSC}
\cosh\mu \cos\nu\,N_{\mu\mu} + \cosh\mu \cos\nu\,N_{\nu\nu}
  - \sinh\mu \cos\nu N_\mu + \cosh\mu \sin\nu N_\nu = 0
\end{equation}
Here, a separation with $\lambda(\lambda-1) \in \C, \Re \lambda>-1$ leads to
\begin{equation}\fl \label{EQ:OSC_sol}
\Psi_\lambda (\mu,\nu) = \cases{ \left[ A_0 \sinh\mu + B_0 \right]\left[ C_0 \sin
\nu + D_0 \right]& $\lambda = 0$ \, , \\
\cos^2\nu \cosh^2\mu \bigl[ A_\lambda P'_\lambda(\rmi\sinh\mu)
 + B_\lambda Q'_\lambda(\rmi\sinh\mu) \bigr]\ast&  \\
\quad \left[ C_\lambda P'_\lambda(\sin\nu) + D_\lambda Q'_\lambda(\sin\nu)
\right]  & $\lambda\neq 0$ \, .}
\end{equation}
where, once again, $P_\lambda$ and $Q_\lambda$ a Legendre functions of the first and second kind.
By assuming integer $\lambda$ and reflection symmetry, \eref{EQ:OSC_sol} can be written as
\begin{equation}\fl \label{EQ:OSC_sol_sym}
\wt{\Psi}_\lambda (\mu,\nu) = \cases{A_0 \sinh\mu + B_0& $\lambda = 0$ \, , \\
 \cos^2\nu \cosh^2\mu P'_{2l+1}(\sin\nu) [ A_{2l+1} P'_{2l+1}(\rmi\sinh\mu) &  \\
 \quad + B_{2l+1} Q'_{2l+1}(\rmi\sinh\mu) ]& $\lambda=2l+1$\, .}
\end{equation}

\end{document}